\documentclass[prl,aps,showpacs,twocolumn,unsortedaddress]{revtex4}
\usepackage{graphics,bm}
\usepackage{amssymb}
\usepackage{epsfig}
\usepackage{epsf}

\begin{document}

\title{Thermally-Assisted Current-Driven Domain Wall Motion}

\author{R.A. Duine}
\email{duine@physics.utexas.edu}
\homepage{http://www.ph.utexas.edu/~duine}

\author{A. S. N\'u\~nez}
\email{alnunez@physics.utexas.edu}
\homepage{http://www.ph.utexas.edu/~alnunez}

\author{A.H. MacDonald}
\email{macd@physics.utexas.edu}
\homepage{http://www.ph.utexas.edu/~macdgrp}

\affiliation{The University of Texas at Austin, Department of
Physics, 1 University Station C1600, Austin, TX 78712-0264}
\date{\today}

\begin{abstract}
Starting from the stochastic Landau-Lifschitz-Gilbert equation, we
derive Langevin equations that describe the nonzero-temperature
dynamics of a rigid domain wall. We derive an expression for the
average drift velocity of the domain wall $\langle \dot{r}_{\rm
dw} \rangle$ as a function of the applied current, and find
qualitative agreement with recent magnetic semiconductor
experiments. Our model implies that at any nonzero temperature
$\langle \dot{r}_{\rm dw} \rangle$ initially varies linearly with
current, even in the absence of non-adiabatic spin torques.
\end{abstract}

\pacs{72.25.Pn, 72.15.Gd}

\maketitle

\def\bx{{\bf x}}
\def\bk{{\bf k}}
\def\half{\frac{1}{2}}
\def\args{(\bx,t)}

\noindent {\it Introduction} --- Possibilities opened up by modern
nanofabrication capabilities have motivated renewed interest in
current-induced domain wall motion, a phenomenon first predicted
and subsequently observed in seminal work by Berger
\cite{berger1984,freitas1985}. The theoretical
\cite{tatara2004,zhang2004,thiaville2004,barnes2005,tserkovnyak2005,kohno2006,
nguyen2006} and experimental
\cite{grollier2003,tsoi2003,yamaguchi2004,klaui2005,yamanouchi2004,yamanouchi2006}
study of domain wall motion induced by spin transfer torques
\cite{slonczewski1996,berger1996,tsoi1998,myers1999} is currently
one of the most active subfields of spintronics \cite{wolf2001}.
Recent research has highlighted a number of fundamentally
interesting issues that are currently under spirited debate. One
controversy concerns intrinsic
pinning\cite{tatara2004,barnes2005}, {\it i.e.}, domain walls that
are stationary up to a critical current in the absence of spatial
inhomogeneity.  An intellectually distinct but phenomenologically
related debate surrounds theories of nonadiabatic spin torques,
which differ widely in their predictions
\cite{zhang2004,thiaville2004,barnes2005,tserkovnyak2005,kohno2006}.
It turns out that in the presence of these torques, a domain wall
is never intrinsically pinned
\cite{zhang2004,thiaville2004,barnes2005}.

This Letter is motivated primarily by the recent experiments of
Yamanouchi {\it et al.} \cite{yamanouchi2006}, in which
current-induced domain wall motion was studied over five orders of
magnitude of average wall velocity.  An important conclusion of
these authors is that at low temperatures the domain wall
undergoes creep motion, {\it i.e.}, that the domain wall does not
move rigidly. Yamanouchi {\it et al.} arrive at this conclusion
because the effects of nonzero temperature, when treated
\cite{tatara2005} in a rigid domain wall approximation, seem to
lead to results that are irreconcilable with experiment. In this
Letter we demonstrate that a systematic theory of the influence of
a thermal bath on the current-driven motion of a rigid domain wall
leads to results that are in qualitative agreement with
experiment.  In the following sections we first explain our theory
of nonzero-temperature domain wall motion and then discuss its
implications for recent experiments.

\noindent {\it Drift velocity of a rigid domain wall} --- Our
starting point is the stochastic Landau-Lifschitz-Gilbert (LLG) equation
\cite{brown1963,kubo1970,ettelaie1984,garciapalacios1998,heinonen2004,safonov2005,rossi2005}
for the direction of magnetization $\hat \Omega$
\begin{equation}
\label{eq:stochasticLLG}
  \frac{\partial \hat \Omega}{\partial t} =\hat \Omega \bm{\times} \left (
  {\bf H} + \bm {\eta} \right) - \alpha \, \hat \Omega
  \bm{\times}  \frac{\partial \hat \Omega}{\partial t}~,
\end{equation}
where ${\bf H}$ is the effective field defined by ${\bf H}(\bx) =
- \delta E_{\rm MM}[\hat{\Omega}]/(\hbar \delta
\hat{\Omega}(\bx))$ and $E_{\rm MM}[\hat{\Omega}]$ is the
micromagnetic energy functional. In Eq.~(\ref{eq:stochasticLLG})
${\bm \eta}$ is a gaussian stochastic magnetic field with zero
mean and correlations
\begin{equation}
\label{eq:noisecorrsfull}
  \langle \eta_\sigma (\bx,t) \eta_{\sigma'} (\bx',t') \rangle
  = \sigma  \delta (t-t') a^3 \delta
  (\bx-\bx') \delta_{\sigma\sigma'}~,
\end{equation}
where $a^3$ is the (local) volume of the finite element grid. The
strength of the noise is given by $\sigma = 2 \alpha k_{\rm B}
T/\hbar$, proportional to the Gilbert damping parameter $\alpha$
and to the thermal energy $k_{\rm B} T$. ($\hbar$ is Planck's
constant which relates energy and frequency.) In using this
expression for the strength of the fluctuations we neglected the
influence of current on thermal magnetization fluctuations
\cite{foros2005} which is higher order \cite{duine2006} than the
spin-torque effects studied here.

The effective field can be separated into magnetic energy and spin transfer
torque contributions ${\bf H}=\left.{\bf H}\right|_{0}+\left.{\bf
H}\right|_j$. For a ferromagnet with an easy $z$ axis and a hard $y$ axis we have
\begin{equation}
\label{eq:efffieldstatic}
 \left. {\bf H} \right|_0 =  J \nabla^2 \hat
 \Omega + 2 \omega_{\rm i}\Omega_z - 2 \omega_{\rm o} \Omega_y~ + H_{\rm ext} \hat{z},
\end{equation}
where $J$ is the spin stiffness, $H_{\rm ext}$ is an external
field in the easy-axis direction and $\omega_{\rm i}$ and
$\omega_{\rm o}$ are respectively the easy axis and hard-plane
anisotropy constants. Assuming only locality implied by smooth
magnetization textures, the spin-transfer torque can be separated
quite generally into contributions parallel and perpendicular to
the spatial derivative of the magnetization:
\begin{equation}
\label{eq:efffieldcurrent}
  \left. {\bf H} \right|_j   \times \hat{\Omega} = v_{\rm s} \frac{\partial \hat \Omega}{\partial r}
   + \beta v_{\rm s} \hat \Omega \bm{\times} \frac{\partial \hat \Omega}{\partial
   r}~,
\end{equation}
where the gradient is taken in the direction of current flow. In
the absence of spin-orbit coupling, it follows from total spin
conservation that $\beta = 0$ and that the {\em spin velocity}
$v_{\rm s} \equiv (j_{\uparrow} - j_{\downarrow})/(-e
(n_{\uparrow} - n_{\downarrow}))$, where $j_{\sigma}$ and
$n_{\sigma}$ are majority and minority spin contributions to the
currents and spin-densities in the collinear limit. For realistic
ferromagnets spin and orbital degrees of freedom are coupled, and
the microscopic theory of $v_s$ and $\beta$ is more challenging
and still controversial
\cite{zhang2004,thiaville2004,barnes2005,tserkovnyak2005,kohno2006}.
The term proportional to $\beta$, the nonadiabatic spin transfer
torque, plays a central role in the theory of current-driven
domain wall motion.

In the absence of current and noise, Eq.~(\ref{eq:stochasticLLG})
admits time-independent solutions corresponding to domain walls.
In terms of the angles $\theta$ and $\phi$ defined by $\hat \Omega
= (\sin \theta \cos \phi, \sin \theta \sin \phi, \cos \theta)$,
the solution corresponding to an isolated domain wall centered at
$r_{\rm dw}$ is $\phi=0$ and $\cos \theta (x-r_{\rm dw}) =\tanh
[(x-r_{\rm dw})/\lambda]$, where the domain wall width
$\lambda=\sqrt{J/(2\omega_{\rm i}})$\,\cite{tatara2004}. Rigid
domain wall motion is described by elevating $r_{\rm dw} \to
r_{\rm dw}(t)$ and $\phi(x,t) \to \phi_0(t)$ to the role of
collective dynamical variables\cite{tatara2004}. The Langevin
equations which describe their stochastic dynamics are
\begin{eqnarray}
\label{eq:langevinvarpars}
    \dot \phi_0 + \alpha \frac{\dot  r_{\rm dw}}{\lambda} &=& \frac{\beta v_{\rm
    s}}{\lambda} -H_{\rm ext} + \eta_{\phi} ~; \\
\label{eq:langevinvarpars2} \frac{\dot r_{\rm dw}}{\lambda} -
\alpha \dot \phi_0 &=&\omega_{\rm o} \sin (2\phi_0) + \frac{v_{\rm
s}}{\lambda} + \eta_r ~.
\end{eqnarray}
These equations can be derived heuristically by enforcing
consistency with the stochastic LLG equations at the center of the
domain wall. Alternately these equations can be derived by noting
that the probability distribution $P[\hat \Omega,t]$, generated by
Eqs.~(\ref{eq:stochasticLLG})~and~(\ref{eq:noisecorrsfull}) can be
written as a path integral $P[\hat \Omega,t] = \int d [\hat
\Omega] \delta [\hat{\Omega} \cdot \hat{\Omega} - 1]
e^{-S[\hat\Omega]}$, with effective action
\cite{duine2006,zinnjustinbook,duine2002}
\begin{equation}
\label{eq:effactionfull}
  S[\hat \Omega] = \int^{t}dt'\int\!\frac{d\bx}{a^3} \frac{1}{2 \sigma}
 \left(
   \frac{\partial \hat \Omega}{\partial t'} \times \hat \Omega - {\bf H} + \alpha
   \frac{\partial \hat \Omega}{\partial t'}
  \right)^2~.
\end{equation}
Inserting the domain-wall solution with time-dependent $r_{\rm
dw}$ and $\phi_0$ into this action gives rigid domain wall
probabilities specified by the effective action
\begin{eqnarray}
\label{eq:effactionsrdwandphi}
 && S[r_{\rm dw},\phi_0] = \int^t dt' \frac{N}{2\sigma}\left[
  \left( \dot \phi_0 + \alpha \frac{\dot  r_{\rm dw}}{\lambda} - \frac{\beta v_{\rm s}}{\lambda}+H_{\rm ext}\right)^2
  \right. \nonumber \\
  && +
  \left( \frac{\dot r_{\rm dw}}{\lambda} - \alpha \dot \phi_0
   -\omega_{\rm o } \sin (2\phi_0) - \frac{v_{\rm s}}{\lambda}\right)^2
  \nonumber \\
   && \left.
   -\frac{4}{3} \omega_{\rm o} ^2 \sin^4 \phi_0
   \right]~,
\end{eqnarray}
where $N=2\lambda A/a^3$ is the number of spins in a domain wall
with cross-sectional area $A$. If we ignore the last term in this
effective action, which vanishes in any case in the gaussian
fluctuation limit,
Eqs.~(\ref{eq:langevinvarpars})~and~(\ref{eq:langevinvarpars2})
are recovered. This minor inconsistency in the theoretical
treatment can be traced to the constant azimuthal angle across the
domain wall in the variational {\em ansatz}. The final term must
also be dropped if the rigid domain wall approximation action is
to reproduce the correct equilibrium probability distribution
function. (Note that even when $\beta=0$ the Boltzmann equilibrium
distribution is approached in the steady state.) The advantage of
this approach is that we can read off the strengths of the
gaussian noise terms $\eta_\phi$ and $\eta_r$ which are given by
\begin{equation}
\label{eq:noiseonvarpars}
 \langle \eta_\phi (t) \eta_\phi (t') \rangle
 = \langle \eta_r (t) \eta_r (t') \rangle
 = \frac{\sigma}{N} \delta (t-t')~.
\end{equation}
Eqs.~(\ref{eq:langevinvarpars}),~(\ref{eq:langevinvarpars2}),~and~(\ref{eq:noiseonvarpars})
generalize the variational approach to current-induced domain-wall
motion of Ref.~\cite{tatara2004} to finite temperature. Tatara and
Kohno \cite{tatara2004} derive their equations from an energy
functional, an approach that has been criticized recently by
Barnes and Maekawa \cite{barnes2005}. Our derivation does not
appeal to an energy functional. Therefore, in addition to deriving
the correct form of the noise terms required for a consistent
treatment of thermal fluctuations, it also provides an alternative
justification of the zero-temperature approach of Tatara and
Kohno.

These equations can be explored numerically without difficulty.
However, it turns out to be possible to make some analytic
progress. Solving for $\dot{\phi_0}$ alone  specializing to purely
current-driven motion ($H_{\rm ext}=0$) we find that
\begin{equation}
\label{eq:langevinphionly}
(1 + \alpha^2) \dot \phi_0 = -  \alpha \omega_{\rm o} \sin (2\phi_0)
 + \frac{(\beta - \alpha) v_{\rm s}}{\lambda}+ \eta_{\phi} - \alpha \eta_{r}~.
\end{equation}
This is the equation of motion for an overdamped Brownian particle
which has been studied extensively with a large number of
different physical motivations.  Our theory for the equation of
motion of the magnetization tilt should be contrasted with the
treatment in Ref.~\cite{tatara2005} which ultimately arrives at an
underdamped Brownian particle equation to describe nonzero
temperature domain wall motion.

The average $\langle \dot \phi_0 \rangle$ can be calculated
exactly for overdamped Brownian motion in a tilted periodic
potential \cite{riskenbook}.  Inserting this well known result in
Eq.~(\ref{eq:langevinvarpars}) we find that
\begin{widetext}
\begin{equation}
\label{eq:fulleqndwvelocity}
   \langle \dot r_{\rm dw} \rangle
\!=\!\frac{\beta v_{\rm s}}{\alpha}-
  \frac{ 2\pi \lambda k_{\rm B} T \left[1\!-\!e^{
       2 \pi \hbar N  (\alpha-\beta) v_{\rm s}
       /(\alpha \lambda k_{\rm B} T) }\right]}
       {\hbar N \left\{\int_0^{2\pi}e^{ V(\phi)/(k_{\rm B} T)}
       d\phi \int_0^{2\pi}e^{- V(\phi')/(k_{\rm B} T)} d\phi'
   \!-\!\left[1\!-\!e^{
       2 \pi \hbar N  (\alpha-\beta) v_{\rm s}
       /(\alpha \lambda k_{\rm B} T) }\right] \int_0^{2\pi}e^{- V(\phi)/(k_{\rm B} T)}
       d\phi \int_0^{\phi}e^{V(\phi')/(k_{\rm B} T)}
       d\phi'\right\}}~,
\end{equation}
\end{widetext}
where
\begin{equation}
\label{eq:potentialphi} V(\phi_0)=- \hbar N \omega_{\rm o} \cos
(2\phi_0)/2+\hbar N(\alpha-\beta)v_{\rm s}\phi_0/(\alpha
\lambda)~,
\end{equation}
is the tilted-washboard potential experienced by the Brownian
``$\phi_0$-particle'' \cite{tatara2004}. In the above equations we
have assumed that $\alpha^2 \ll 1$, for notational convenience.

It is well-known \cite{riskenbook} that at zero temperature the
equation for an overdamped particle in the tilted periodic
potential of Eq.~(\ref{eq:langevinphionly}) has solutions with
$\langle \dot \phi \rangle \neq 0$ only if $|(\alpha - \beta )
v_{\rm s} |> \alpha \lambda \omega_{\rm o}$. For $\beta=0$ the
zero temperature result for the average velocity of the domain
wall is $\langle \dot r_{\rm dw} \rangle \propto \sqrt{(v_{\rm
s}/v_{\rm sc})^2-1}$ \cite{tatara2004,riskenbook}, where the
critical current for depinning the domain wall is $v_{\rm sc} =
\lambda \omega_{\rm o}$ \cite{tatara2004}.  For $\beta \ne 0$, the
average domain wall velocity is nonzero at any finite value of
$v_s$ even at zero temperature.

\begin{figure}
\vspace{-0.5cm} \centerline{\epsfig{figure=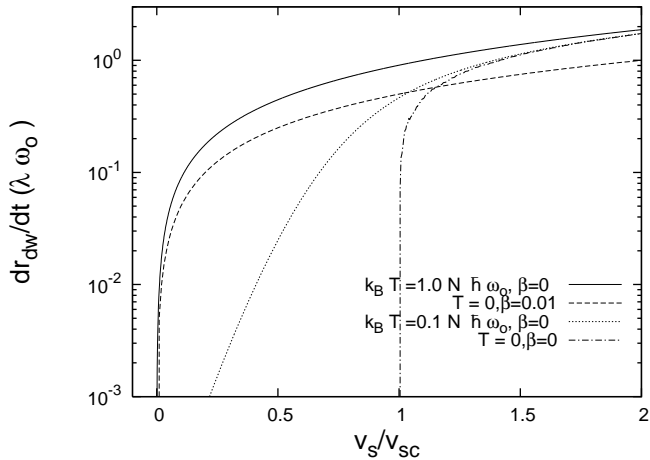}}
 \caption{Domain wall velocity as a function of current,
 for various temperatures and values of $\beta$. We take $\alpha=0.02$.
 The horizontal axis is normalized to $v_{\rm sc}=\lambda \omega_{\rm o}$.}
 \label{fig:v_j}
\end{figure}

In Fig.~\ref{fig:v_j} we plot the average domain-wall velocity
calculated from the full expression
[Eq.~(\ref{eq:fulleqndwvelocity})] for various temperatures and
values of $\beta$. These curves were calculated for Gilbert
damping parameter $\alpha=0.02$. We consider only $\beta<\alpha$
since for $\beta>\alpha$ the domain wall velocity would decrease
with increasing temperature in clear disagreement with experiment
\cite{yamanouchi2006}. One of the main conclusions of
Ref.~\cite{tatara2005} is that for $\beta = 0$ and small currents
the average domain-wall velocity $\langle \dot r_{\rm dw} \rangle
\propto \exp(C v_{\rm s})$ where $C$ is a constant. In the
low-temperature limit we find that for $v_{\rm s}/v_{\rm sc} < 1$
the average domain wall velocity is given by
\begin{widetext}
\begin{equation}
  \langle
  \dot r_{\rm dw} \rangle
   = \frac{\beta v_{\rm s}}{\alpha}
   + 2 \sqrt{\left(\lambda \omega_{\rm o}\right)^2 - \left( v_{\rm s}\right)^2}
   \exp\left\{ \left(\frac{\beta}{\alpha} -1 \right)
   \frac{N \hbar \omega_{\rm o}}{k_{\rm B} T}\left[
    \sqrt{1-\left( \frac{v_{\rm s}}{v_{\rm sc}}\right)^2} + \left(\frac{v_{\rm s}}{v_{\rm
    sc}}\right)
    \sin^{-1} \left(\frac{v_{\rm s}}{v_{\rm sc}} \right)\right]\right\}
  \sinh \left( \frac{\pi \hbar N \left( \alpha - \beta \right) v_{\rm s}}{\alpha \lambda k_{\rm B} T}
\right)~.
\label{lowT}
\end{equation}
\end{widetext}
This difference in estimated domain wall velocities is closely
analogous to the difference between the Langer-Ambegaokar
\cite{LA} and Halperin-McCumber \cite{HMcC} estimates of phase
slip resistance in thin superconducting wires. It follows from
Eq.~(\ref{lowT}) that the observation of linear dependence of
domain wall velocity on current does not necessarily imply that
$\beta \neq 0$. A careful analysis of the temperature dependence
of $\langle \dot r_{\rm dw} \rangle$ will be necessary to
determine the value of $\beta$ from low current experiments,
especially so if $N \hbar \omega_{\rm o}$ is not very large
compared to $k_{\rm B} T$. Because of thermal noise, even when
$\beta=0$ the domain is not intrinsically pinned at any nonzero
temperature.

\noindent {\it Comparison with experiment} --- The results
presented in Fig.~\ref{fig:v_j} look qualitatively  similar to the
experimental results of Yamanouchi {\it et al.}
\cite{yamanouchi2006}, in particular to the inset in Fig.~3 of
Ref.~\cite{yamanouchi2006}. A direct comparison with these
experimental results is complicated by the fact that they are
performed close to the critical temperature, with the result that
the magnetic anisotropy energy-density and the polarization of the
current depend on temperature. Moreover, the Gilbert-damping
parameter $\alpha$ may also depend on temperature. The fact that
curves at different temperatures in Fig.~\ref{fig:v_j} are not as
strongly offset vertically  from each other at $v_{\rm s}/v_{\rm
sc}>1$ as the experimental results is most likely due to these
additional temperature-dependent effects.

\begin{figure}
\vspace{-0.5cm} \centerline{\epsfig{figure=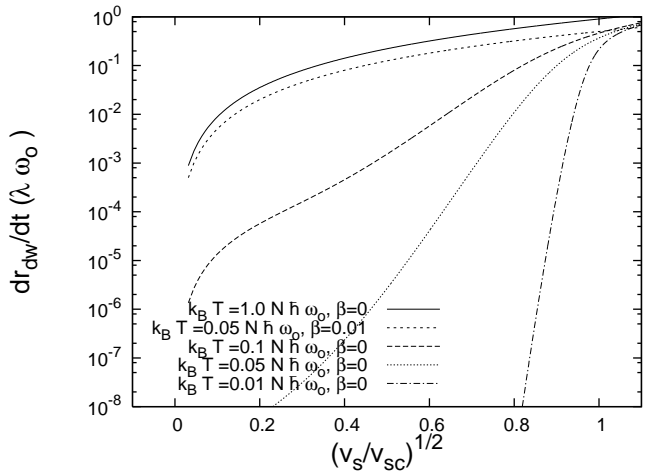}}
 \caption{Test of the scaling $\ln \langle \dot r_{\rm dw} \rangle \propto \sqrt{v_{\rm s}}$,
 for various temperatures and values of $\beta$. We take
 $\alpha=0.02$.}
 \label{fig:v_sqrtj}
\end{figure}

One of the main results of Yamanouchi {\it et al.} is the
empirical finding that for small currents $\ln \langle \dot r_{\rm
dw} \rangle \propto \sqrt{ v_{\rm s}}$. From this the authors
conclude that the domain wall undergoes current-induced creep
motion at small currents. In Fig.~\ref{fig:v_sqrtj} we plot the
average domain wall velocity as a function of $\sqrt{v_{\rm s}}$
for various temperatures and values of $\beta$. From this we
observe that $\ln \langle \dot r_{\rm dw} \rangle$ is proportional
to the square root of the current over several decades of domain
wall velocity. It turns out that this approximate relationship is
valid for a larger range of parameter space with decreasing
temperature. In the experiments of Yamanouchi {\it et al.} a
typical velocity is $\langle \dot r_{\rm dw} \rangle \sim 1$ m
s$^{-1}$. Putting this equal to $\lambda \omega_{\rm o}$ we have
for $\lambda = 17$ nm and $T \sim 100$ K that $k_{\rm B} T/(N
\hbar \omega_{\rm o}) \sim 0.01$ for $N=3.4\times10^6$ Mn moments
(we take the density of moments $N_{\rm Mn}\sim 1$ nm$^{-3}$) in a
wall of cross-sectional area $A=2$ nm $\times$ $5$ $\mu$m. For
this value of $k_{\rm B} T/(N \hbar \omega_{\rm o})$ we conclude
from Fig.~\ref{fig:v_sqrtj} that $\ln \langle \dot r_{\rm dw}
\rangle \propto \sqrt{v_{\rm s}}$ over a large regime, and that
our theory is therefore in agreement with the experimental results
of Yamanouchi {\it et al.} \cite{yamanouchi2006}. Moreover, from
Fig.~\ref{fig:v_sqrtj} we also conclude that for $\beta=0.01$ this
scaling does not hold. Hence, an additional conclusion from the
experimental data is that $\beta$ is much smaller than  $\alpha$
in these ferromagnetic semiconductors.

A similar estimate for typical parameters of the metallic
nanowires used in studies of current-driven domain wall motion
\cite{grollier2003,tsoi2003,yamaguchi2004,klaui2005} leads to the
conclusion that $k_{\rm B} T/(N\hbar\omega_{\rm o})\sim0.0002$ and
therefore that temperature effects on rigid domain wall motion are
likely much less important in these systems. This difference
arises mainly because the density of moments is roughly $40$ times
higher in metallic systems.

In conclusion, we have presented a theory of the influence of
nonzero temperatures on current-driven motion of a rigid domain
wall, and found qualitative agreement with ferromagnetic
semiconductor experiments. An estimate of the bending energy of
the domain wall shows that these degrees of freedom are also
thermally accessible. It is, however, difficult to assess how they
influence translation of the domain wall. The qualitative
agreement of our results with the experiments of Yamanouchi {\it
et al.} \cite{yamanouchi2006} indicates that thermal activation of
rigid domain wall motion could play an important role in these
experiments. We expect that accounting for thermal fluctuations
will also be important in assessing the impact of intended and
unintended extrinsic domain wall pinning.   This work was
supported by the National Science Foundation under grants
DMR-0115947 and DMR-0210383, by a grant from Seagate Corporation,
and by the Welch Foundation.

\end{document}